\documentclass[aps,prc,amsmath,showpacs,superscriptaddress,nofootinbib]{revtex4}
\usepackage{graphicx,epsfig,epstopdf,longtable}
\newcommand{\beqy}{\begin{eqnarray}}
\newcommand{\eeqy}{\end{eqnarray}}
\newcommand{\bmlet}{\begin{subequations}}
\newcommand{\emlet}{\end{subequations}}

\newcommand{\Mns}{\mathcal{M}}
\newcommand{\Rns}{R_\star}

\usepackage[normalem]{ulem}
\usepackage[dvipsnames]{xcolor}

\begin{document}

\def\gsimeq{\,\,\raise0.14em\hbox{$>$}\kern-0.76em\lower0.28em\hbox
{$\sim$}\,\,}
\def\lsimeq{\,\,\raise0.14em\hbox{$<$}\kern-0.76em\lower0.28em\hbox
{$\sim$}\,\,}

\title{Unified equations of state for cold non-accreting neutron stars with 
Brussels-Montreal functionals.  IV. Role of the symmetry energy in pasta phases} 

\author{N. N. Shchechilin}
\affiliation{Institut d'Astronomie et d'Astrophysique, CP-226, Universit\'e
	Libre de Bruxelles, 1050 Brussels, Belgium}

\author{N. Chamel}
\affiliation{Institut d'Astronomie et d'Astrophysique, CP-226, Universit\'e
Libre de Bruxelles, 1050 Brussels, Belgium}

\author{J. M. Pearson}
\affiliation{D\'ept. de Physique, Universit\'e de Montr\'eal, Montr\'eal
(Qu\'ebec), H3C 3J7 Canada}

\date{\today}

\begin{abstract}
Our previous investigation of neutron-star crusts, based on the functional BSk24,  led to a substantial reduction of the pasta mantle when Strutinsky integral and pairing corrections were added on top of the fourth-order extended Thomas-Fermi method (ETF). 
Here, our earlier calculations are widened to a larger set of functionals within the same family, and we find that the microscopic corrections weaken significantly the influence of the symmetry energy. 
In particular, the correlation observed at the pure ETF level between the density for the onset of pasta formation and the symmetry energy vanishes, not only for the $L$ coefficient but also for the symmetry-energy values at the relevant densities. Moreover, the inclusion of microscopic corrections results in a much lower abundance of pasta for all functionals. 
\end{abstract}

\maketitle

\section{Introduction}
\label{intro}

In 2018 we published unified calculations of the equation of state (EoS) and
the composition of all regions of a cold non-accreting neutron star, with the 
same highly realistic nuclear energy-density functionals being used in all regions of the star, i.e., the outer crust, the inner crust and the core~\cite{Pearson_ea18_bsk22-26}. The inner crust was calculated in the framework of spherical Wigner-Seitz (WS) cells using the  ETFSI (fourth-order 
extended Thomas-Fermi plus Strutinsky integral) method~\cite{dutta2004,Onsi+08,Pearson+12,Pearson_ea15_pairing}.
This is a high-speed approximation to the Hartree-Fock-Bogoliubov (HFB) method 
(for a review of which see, for example, Ref.~\cite{Bender+2003}), consisting of two 
distinct stages: a full extended Thomas-Fermi (ETF) treatment of the kinetic-energy and spin-current 
densities, followed by the addition of proton shell corrections, calculated by 
the Strutinsky-integral method (SI), and proton pairing correlations, handled in
the Bardeen-Cooper-Schrieffer (BCS) approximation (see, e.g., Ref.~\cite{Shelley_Pastore20}
for recent comparisons between the HFB and ETFSI methods in neutron-star crusts). 

More recently, we have extended our calculations of the inner crust to admit the 
possibility of nuclear ``pasta'' phases considering infinitely long rods 
(``spaghetti'') and tubes (``bucatini''), plates of infinite extent 
(``lasagna'') and spherical bubbles (``Swiss cheese''). First predicted by 
Ravenhall {\it et al.}~\cite{Ravenhall_ea83} 
and Hashimoto {\it et al.}~\cite{Hashimoto+84} within the liquid-drop approach, 
nuclear pasta could represent more than $50\%$ of the mass of the crust of neutron 
stars~\cite{Lorenz+93}, as supported by recent liquid-drop model calculations 
(see, e.g., Refs.~\cite{DinhThi+21a,Balliet+21,Parmar+22}). Nuclear pasta phases have 
also been found with more realistic semi-classical treatments
(see, e.g., Refs.~\cite{Oyamatsu93,Okamoto+13,Oyamatsu&Iida07,Grill+13,Bao&Shen15,MU15,JiHuShen21,Sharma_Centelles+15}), and this has been confirmed by our ETF calculations~\cite{Pearson+20}. 
A striking conclusion of Ref.~\cite{Pearson+22} was that when the SI and pairing
corrections for protons were included in the calculations the spaghetti phase 
vanished completely, with the spherical clusters (sometimes referred to as 
``gnocchi'' or ``polpete'') dominating everywhere except for a thin layer of 
lasagna close to the interface with the homogeneous core. In our calculations, 
bucatini and Swiss cheese were allowed but we did not find any stable configurations. 

The calculations of Refs.~\cite{Pearson+20,Pearson+22} were performed with the 
nuclear energy-density functional BSk24, a member of an extended family of 
functionals that were developed not only for the study of neutron-star 
structure but also for the general purpose of providing a unified treatment of
a wide variety of phenomena associated with the birth and death of a neutron 
star, such as core-collapse supernova and neutron-star mergers, along with 
the r-process of nucleosynthesis (see Ref.~\cite{Goriely_ea_Bsk22-26} and references 
therein). These functionals are based on generalized Skyrme-type forces and 
density-dependent contact pairing forces, the formalism for which is presented
in Refs.~\cite{Chamel+09,Goriely+09_pair,Chamel2010}. The parameters of the functional 
were determined primarily by fitting to essentially all the nuclear-mass data of 
the 2012 Atomic Mass Evaluation~\cite{ame12}; nuclear masses were calculated 
using the HFB method, with axial deformation taken into account. In making 
these fits certain constraints were imposed, the most significant of which is to
require consistency, up to the densities prevailing in neutron-star cores, 
with the EoS of homogeneous pure neutron matter, as calculated {\it ab initio}
from realistic two- and three-nucleon forces. The resulting unified EoS was found 
to be consistent with constraints inferred from astrophysical observations of 
neutron stars, including LIGO-Virgo data on GW170817~\cite{Perot2019,Perot2022}. 

Actually, unified EoSs based on the other functionals BSk22, BSk25, and BSk26 of the 
same family were also calculated in Ref.~\cite{Pearson_ea18_bsk22-26}. The functionals 
BSk22 and BSk25 were adjusted to different values of the symmetry-energy coefficient $J$, 
whereas BSk26 differs in the choice of the realistic neutron-matter EoS. Here $J$ is defined 
as 
\beqy 
J\equiv \frac{1}{2}\dfrac{\partial^2 e(n_0,\eta)}{\partial\eta^2}\biggr\vert_{\eta=0} \, , 
\eeqy 
where $e(n,\eta)$ stands for the energy per nucleon of infinite homogeneous nuclear matter 
at density $n$ and isospin asymmetry $\eta=(n_n-n_p)/n$ ($n_n$ and $n_p$ are the neutron 
and proton densities respectively), and $n_0$ is the equilibrium density of symmetric 
nuclear matter. The properties of asymmetric nuclear matter can be further characterized 
by the higher-order coefficients $L$ and $K_{\rm sym}$ appearing in the expansion of the 
energy per nucleon up to second order in the isospin asymmetry $\eta$ and in the dimensionless parameter 
$\epsilon=(n-n_0)/n$:
\beqy 
\label{eq:e-expansion}
e(n,\eta)\approx e(n_0,0) + \frac{1}{18} K_v \epsilon^2 + (J + \frac{1}{3} L \epsilon + \frac{1}{18} K_{\rm sym} \epsilon^2) \eta^2 + ...\, ,
\eeqy 

In the region of the inner crust of neutron stars where pasta phases might appear, the inhomogeneous matter is 
extremely neutron-rich $\eta\sim 0.9$ and the average baryon density $\bar n$ generally lies 
below $n_0/2$, therefore the neglected higher-order terms in the expansion (\ref{eq:e-expansion}) could become important. It is therefore more insightful  to introduce the symmetry energy 
\beqy \label{eq:symmetry-energy definition}
S(n)\equiv e(n,1)-e(n,0)\, ,
\eeqy 
and expand it as
\beqy\label{eq:symmetry-energy-def}
S(n)\approx \widetilde{J} + \frac{1}{3} \widetilde{L} \epsilon + \frac{1}{18} \widetilde{K}_{\rm sym} \epsilon^2 \, , 
\eeqy 
where 
\beqy 
\widetilde{J}\equiv S(n_0) \, , 
\eeqy 
the slope  
\beqy 
\widetilde{L}\equiv 3 n_0 \dfrac{dS}{dn}\biggr\vert_{n=n_0} \, , 
\eeqy 
and 
\beqy 
\widetilde{K}_{\rm sym}\equiv 9 n_0^2 \frac{d^2S}{dn^2}\biggr\vert_{n=n_0} \,  .
\eeqy 

\begin{figure}
 	\includegraphics[width=0.6\columnwidth]{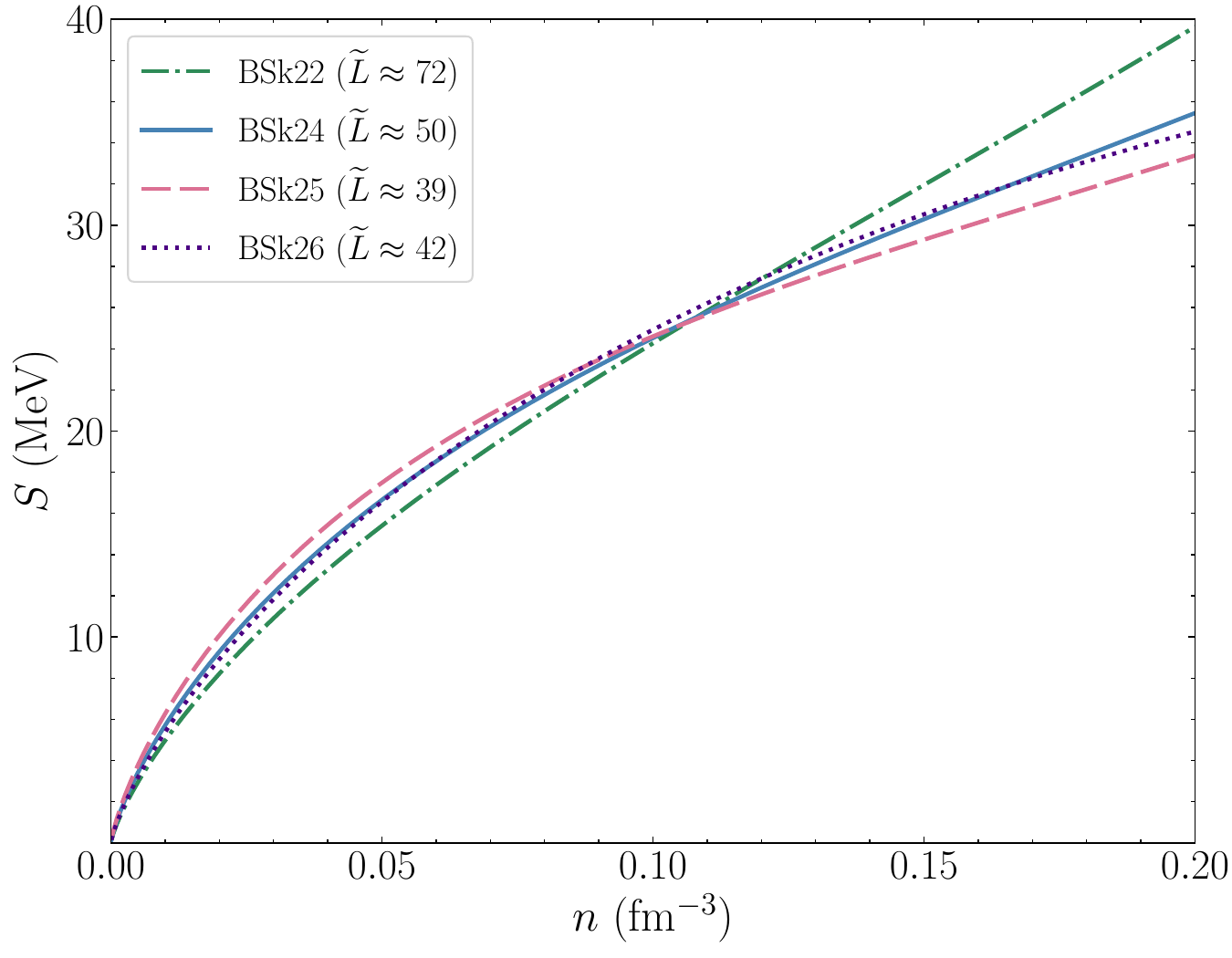}
 	\caption{Symmetry energy~\eqref{eq:symmetry-energy definition} (in MeV) as a function of the baryon density $n$ (in fm$^{-3}$) for the following Brussels-Montreal nuclear energy-density functionals: BSk22 (green dash-dotted), BSk24 (blue solid), BSk25 (red dashed), BSk26 (purple dotted).}
 	\label{fig:symm}
 \end{figure}

The values of all these coefficients for the different BSk functionals are indicated in Table~\ref{tab1}. 
These coefficients are not completely independent. In particular, the fit to nuclear masses 
leads to a correlation between $J$ and $L$ (see, e.g., Ref.~\cite{Goriely&Capote14_JL}) and fixes 
the value of $S$ at the baryon density $n\approx0.11$\,fm$^{-3}$, as can be seen in Fig.~\ref{fig:symm} 
(see also Ref.~\cite{Zhang&Chen_sym13}). Moreover, a higher value of $\widetilde L$ (or $L$) translates into a lower symmetry
energy $S(n)$ at densities $n\lesssim0.11$\,fm$^{-3}$, as can be seen by comparing the predictions from BSk22, 
BSk24, and BSk25 in Fig.~\ref{fig:symm}. Being fitted to a softer neutron-matter EoS with the same $J$ as for 
BSk24, the functional BSk26 yields a softer symmetry energy at densities $n\gtrsim0.11$\,fm$^{-3}$, as reflected 
in the lower values for $\widetilde L$ and $\widetilde K_{\rm sym}$. As shown in Ref.~\cite{Pearson_ea18_bsk22-26} within the ETFSI 
approach, the different behaviors of the symmetry energy $S(n)$ can significantly change the equilibrium composition 
of the inner crust, especially in the densest layers. However, only spherical clusters were considered. 

 \begin{table}
 	\centering
 	\caption{Symmetry-energy coefficients of different BSk functionals. See text for definitions.} 
 \label{tab1}
 \begin{tabular} {ccccc}
 	\hline 
 	&BSk22&BSk24&BSk25&BSk26\\
	\hline
 	$J$\,[MeV]&32.0&30.0&29.0&30.0 \\
 	$\widetilde{J}$\,[MeV]&33.1&31.1&30.0&31.3 \\
 	$L$\,[MeV]&68.5&46.4&36.9&37.5 \\
 	$\widetilde{L}$\,[MeV]&71.7&49.5&39.2&42.2 \\
 	$K_\mathrm{sym}$\,[MeV]&13.0&-37.6&-28.5&-135.6 \\
  	$\widetilde{K}_\mathrm{sym}$\,[MeV]&12.6&-38.2&-32.7&-130.3 \\
 	\hline
 \end{tabular}
\end{table}

The correlation between the region of neutron stars containing nuclear pasta and the symmetry energy has been previously studied within the liquid-drop picture~\cite{Newton+13, Balliet+21,DinhThi+21a,Parmar+22} and the Thomas-Fermi (TF) method~\cite{Oyamatsu&Iida07,Bao&Shen15,Grill+13,JiHuShen21}. Despite the variety of models employed, reducing the slope $L$ has been found to increase the abundance of nuclear pasta. However, important microscopic aspects such as shell structure and pairing are not taken into account in such (semi)classical models. On the other hand, a fully microscopic treatment remains computationally extremely costly. For this reason, pairing is often 
neglected and calculations are restricted to fixed proton fractions rather than $\beta$-equilibrium, as in Ref.~\cite{Fattoyev+17}. In this paper, we pursue our study 
of neutron-star interiors within the ETFSI approach considering the BSk22, BSk25 and BSk26 functionals to better understand the role of the symmetry energy in the formation of nuclear pasta. 

 Our formalism is briefly reviewed in Section~\ref{sec:form}, while results are presented and discussed in Section~\ref{sec:res}. The main conclusions are outlined in Section~\ref{sec:concl}.

\section{Formalism}
\label{sec:form}

\subsection{Extended Thomas-Fermi approach}

We consider five different nuclear pasta phases with WS cells having 3 
different geometries: 
spherical cells for gnocchi and Swiss cheese, infinitely long cylinders for spaghetti and bucatini, and plates of infinite extent for lasagna. We start from the ETF approximation~\cite{Brack_ea85} and minimize the energy per nucleon $e_{\mathrm{ETF}}$ at fixed baryon number density $\bar n$ to determine the equilibrium configuration. More specifically, the energy per nucleon can be written as
\beqy\label{eq:eetf}
e_{\mathrm{ETF}} = e_\mathrm{Sky} + e_\mathrm{C} + e_{e} - Y_{p}\,Q_{n,\beta} \quad ,
\eeqy
where the first term corresponds to the energy of the generalized Skyrme force, the second term accounts for the Coulomb interactions, while the third term is the kinetic energy per nucleon of a relativistic electron Fermi gas (see Refs.~\cite{Pearson_ea18_bsk22-26,Pearson+20} for details). As in our previous studies, the neutron mass is omitted for convenience, $Y_p$ denotes the proton fraction and $Q_{n,\beta}$ represents the neutron beta-decay energy. The contribution from the generalized Skyrme interaction is given by a functional of the local nucleon number densities $n_q(\pmb{r})$, kinetic energy densities  
$\tau_q(\pmb{r})$ and spin-current densities $\pmb{J_q}(\pmb{r})$ at position $\pmb{r}$ ($q=n,p$ for neutrons, protons respectively) of the following form: 
\beqy\label{eq:enuc}
e_\mathrm{Sky} = \frac{1}{A}\int_{\mathrm{cell}}
{\mathcal E_\mathrm{Sky}}(\pmb{r})\,d^3\pmb{r} \quad .
\eeqy  
Here $A$ is the total number of nucleons in the WS cell for the spherical case, the
number per unit length for cylindrical cells (the integration being taken over unit length), or the number per unit area for plate-like cells (the integration being taken over
unit area). The expression of ${\mathcal E_\mathrm{Sky}}$ in terms of the various densities and currents including both density- and momentum-dependent terms can be found in Appendix of Ref.~\cite{Chamel+09}. 
We take into account the finite size of protons as described in 
Ref.~\cite{Pearson+91}, namely, we add an extra Skyrme type $t_0$ interaction term for protons with a strength of -4.8 MeV corresponding to a root-mean-square charge radius of 0.8 fm for the proton. 
This feature has been included in all our subsequent ETF(SI) calculations, although the papers failed to mention it~\cite{Pearson+12,Pearson_ea15_pairing,Pearson_ea18_bsk22-26,Pearson+20,Pearson+22}.
We make use of the full 4th order ETF method to expand $\tau_q(\pmb{r})$  and $J_q(\pmb{r})$ in terms of the  densities $n_q(\pmb{r})$ and their derivatives~\cite{Onsi+08}. Rather than perform full Euler-Lagrange
calculations we speed up the calculations by parametrizing the nucleon 
distributions in the following way:
\beqy\label{eq:rhoq}
{n_q}(\xi) = n_{\mathrm{B}q} + n_{\Lambda q}f_q(\xi)  \quad ,
\eeqy
where the first and second terms represent the uniform nucleonic background and the clusters respectively, with the function given by
\beqy\label{eq:fq}
f_q(\xi) = \frac{1}{1 + \exp \left[\Big(\frac{C_q - R}
	{\xi - R}\Big)^2 - 1\right] \exp \Big(\frac{\xi-C_q}{a_q}\Big) }\, .
\eeqy
Here $\xi$ represents the radial coordinate for spheres and cylinders and the coordinate perpendicular to the plane for plates. $C_q$ and $a_q$ stand for the cluster radius and surface diffuseness respectively, while $R$ is the radius of the spherical WS cell, the radius of the cylindrical cell, or the semi-thickness of the plate-like cell. Note that inverted configurations (tubes and bubbles) correspond to  $n_{\Lambda q}<0$. The advantage of this parametrization is the vanishing of all the derivatives at the border of the cell, which allows for simplifications of the ETF energy using integration by parts. 

\subsection{Extended Thomas-Fermi plus Strutinsky integral approach}
 
In a second stage, we add microscopic corrections for protons as in Ref.~\cite{Pearson+22}:  
 \beqy\label{eq:eetfsi}
 e_\mathrm{ETFSI} = e_\mathrm{ETF} + 
 \frac{1}{A}\left(E^\mathrm{SI}_p + E_p^\mathrm{pair}\right)  \quad  .
 \eeqy
The proton pairing energy $E_p^\mathrm{pair}$ is given by 
\begin{equation}
E_p^\mathrm{pair} =-\frac{3}{8}\int d^3\pmb{r}\, 
n_p(\pmb{r})\frac{\Delta_p(\pmb{r})^2}{\epsilon_F(\pmb{r})} \quad ,
\end{equation}
where $\Delta_p(\pmb{r})$ is the proton pairing gap of homogeneous nuclear
matter of the appropriate local density and charge asymmetry. Also, $\epsilon_F(\pmb{r})$ is
the local proton Fermi energy,
\begin{equation}
\epsilon_F(\pmb{r}) = \frac{\hbar^2}{2M_p}\left(3\pi^2n_p(\pmb{r})\right)^{2/3} \quad .
\end{equation}
The term $E^\mathrm{SI}_p$ is calculated within the Strutinsky integral method, for which we have to solve the Schr\"odinger equation  
\begin{equation}\label{eq:HF}
	\left\{-\pmb{\nabla}\frac{\hbar^2}{2\widetilde{M^*_p}(\pmb{r})}\cdot\pmb{\nabla} +
	\widetilde{U_p}(\pmb{r}) + \widetilde{U_\mathrm{C}}(\pmb{r}) - 
	i\widetilde{\pmb{W_p}}(\pmb{r})\cdot\pmb{\nabla}
	\times\pmb{\sigma}\right\}\psi_{\nu,p}=
	\widetilde{\epsilon}_{\nu,p}\psi_{\nu,p} \quad ,
\end{equation}
where $\psi_{\nu,p}(\pmb{r})$ is the proton wave function associated with the quantum state $\nu$ 
and $\widetilde{\epsilon}_{\nu,p}$ the associated single-particle energy. Here $\pmb{\sigma}$ denotes
the Pauli spin matrices. 
The proton effective mass $\widetilde{M^*_p}(\pmb{r})$, the proton scalar potential $\widetilde{U_p}(\pmb{r})$ and the proton spin-orbit vector potential  $\widetilde{\pmb{W_p}}(\pmb{r})$ are calculated from Eqs.~(A10), (A11) and (A12) of Ref.~\cite{Chamel+09} respectively using the smooth ETF nucleon densities $\widetilde{n_q}(\pmb{r})$ and
corresponding ETF kinetic densities  $\widetilde{\tau_q}(\pmb{r})$ and spin current densities $\widetilde{\pmb{J_q}}(\pmb{r})$. The Coulomb potential $\widetilde{U_\mathrm{C}}(\pmb{r})$ is 
obtained from the solution of Poisson's equation inside the WS cell (see Appendix of Ref.~\cite{Pearson+22}). 
The correction then reads~\cite{Pearson+12}
\begin{equation}\label{eq:SI-correction}
	E^\mathrm{SI}_p = \sum\limits_{\nu}^{\rm occ}\,\widetilde{\epsilon}_{\nu,p} -
	\int d^3\pmb{r}\Biggl\{\frac{\hbar^2}{2\widetilde{M^*_p}(\pmb{r})}
	\widetilde{\tau_p}(\pmb{r}) + \widetilde{n_p}(\pmb{r})\left[\widetilde{U_p}(\pmb{r})
	+\widetilde{U_\mathrm{C}}(\pmb{r}) \right]
	+\widetilde{\pmb{J_p}}(\pmb{r})\cdot\widetilde{\pmb{W_p}}(\pmb{r})\Bigg\}
	\quad ,
\end{equation}
where the summation only goes over occupied states. 
In the region where pasta phases are likely to appear, the spin-orbit coupling in 
Eq.~\eqref{eq:HF} is very small and therefore we drop it. The validity of this approximation is simply a consequence of the near-homogeneity of the nucleonic distributions at those densities (see Section IIIB of Ref.~\cite{Pearson+22}). The first term in 
Eq.~\eqref{eq:SI-correction} involves integrations over wavevectors associated with
longitudinal motions along the symmetry axis for spaghetti and in the symmetry plane
for lasagna. To further simplify the calculations, we set the proton effective mass $\widetilde{M^*_p}(\pmb{r})$ equal to the real proton mass $M_p$ and we correct for this substitution by adding a term to the potential $\widetilde{U_p}(\pmb{r})$, as described 
in Section IIIB of Ref.~\cite{Pearson+22}. In this way, integrations over wavevectors can then be performed analytically (see
Ref.~\cite{Pearson+22} for details). For consistency, spherical configurations are treated using the same approximations. 

In the deepest regions of the crust, our Fortran code applied in our previous studies failed to converge for some densities. To improve the reliability of our calculations, we have rewritten our code in Python making use of the \texttt{SciPy} library for the minimization procedure.

\section{Results}
\label{sec:res}

\subsection{Extended Thomas-Fermi predictions}
 \label{sec:semi}
 
For the given average baryon density $\bar n$, we first calculate the configurations  
yielding the lowest ETF energy per nucleon for each type of pasta. That is to 
say, the energy per nucleon~\eqref{eq:eetf} is minimized with respect to the 
geometrical parameters of the WS cell, as defined in Eqs.~\eqref{eq:rhoq} and \eqref{eq:fq}. 
We have checked that 
the results obtained for spherical clusters, spaghetti and lasagna are in excellent agreement with 
those reported earlier with the functional BSk24~\cite{Pearson+20,Pearson+22}. However,
our new Python code has allowed us to refine the solutions for bucatini and 
Swiss cheese \footnote{In fact, previously within the Fortran program, we missed such inverse configurations due to the too restricted range for the parameters we had to set to stabilize the computations. Now, using an initial guess from calculations in Python, we obtain the same solutions in Fortran code, thus confirming our results for bucatini and Swiss cheese.}. 
Comparing the energy per nucleon between all five pasta phases, we have determined the most
energetically favorable (globally stable) one. 
 
\begin{figure}
\includegraphics[width=0.7\columnwidth,angle=-0]{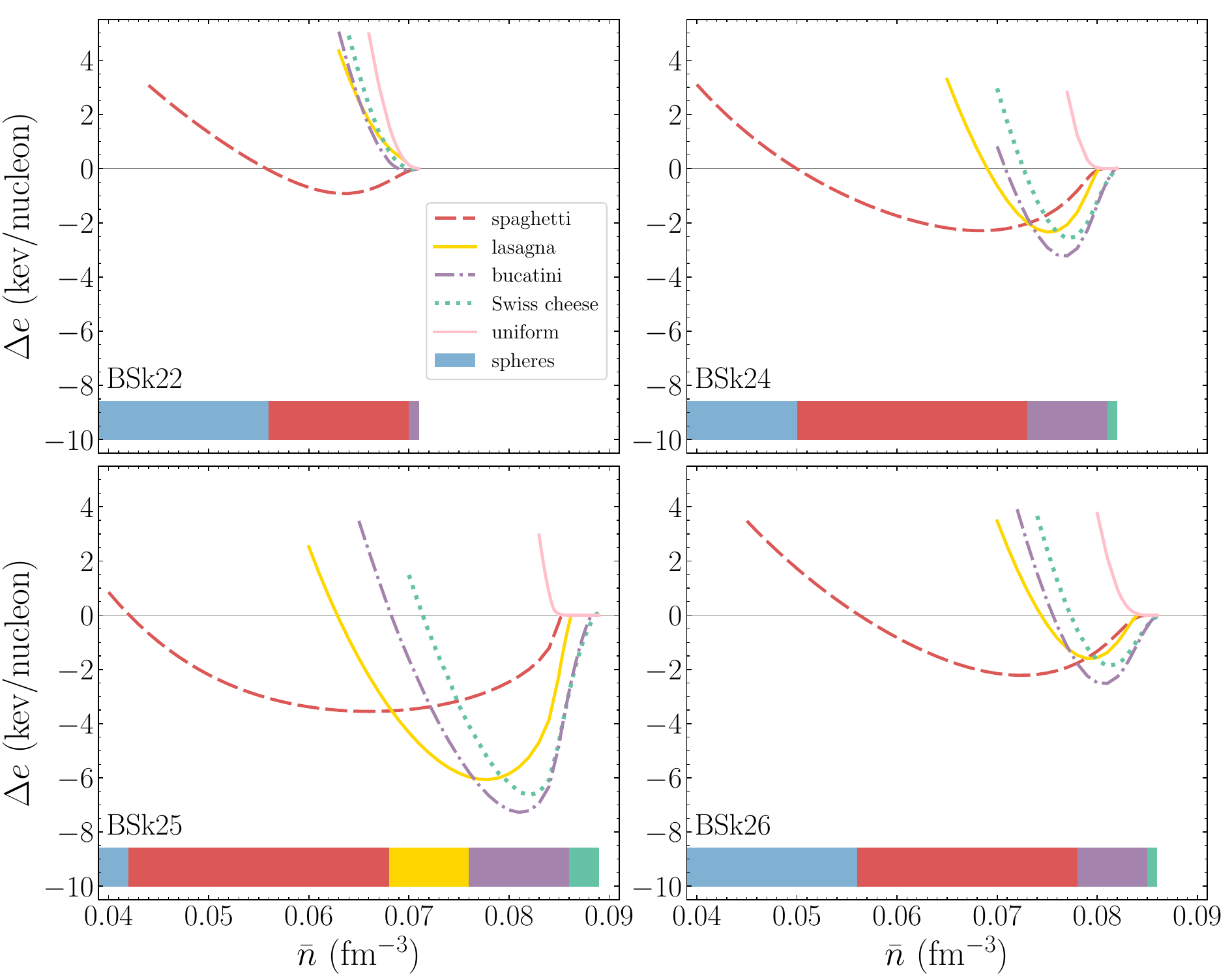}
\caption{ETF energy per nucleon $e$ (in MeV) for different pasta shapes with the energy per nucleon $e_0$ of spheres subtracted versus average baryon number density $\bar n$ (in fm$^{-3}$). Each of the four panels shows results for a given BSk functional belonging to the same family but differing in the prediction 
for the symmetry energy defined by Eq.~\eqref{eq:symmetry-energy-def}. The energies of spaghetti, lasagna, bucatini, Swiss cheese and homogeneous matter are represented by red dashed lines, yellow solid lines, purple dash-dotted lines, green dotted lines and pink solid lines respectively. The horizontal histograms illustrate the sequence of pasta phases with the same color coding, adding a blue area for spheres.}
\label{fig:etf}
\end{figure}

Our results are presented in Figure~\ref{fig:etf}. The ETF energy per nucleon for spheres
is subtracted from each phase for convenience. Within the ETF approach, the onset of pasta phases is characterized by the formation of spaghetti for all adopted functionals. In contrast to our previous results for
BSk24~\cite{Pearson+20,Pearson+22}, lasagna is now replaced by bucatini and Swiss cheese.  
Comparing results
obtained for BSk22 and BSk25 (in Figure \ref{fig:etf} and Table \ref{tab2}), we find that the symmetry energy plays some role in the
existence of pasta phases. Whereas BSk22 only allows for cylindrical shapes, all ``traditional'' pasta phases are present for BSk25 filling a much wider density range. Therefore higher values of the symmetry energy at the relevant densities (corresponding to the lower $\widetilde J$ and $\widetilde{L}$, 
see also Figure \ref{fig:symm}) thus tend to favor pasta. It should also be noticed that the transition
density $\bar n_\mathrm{sp}$ from spheres to pasta is the highest for BSk22 and the lowest
for BSk25. This result is consistent with 
the weak correlation between $\bar n_\mathrm{sp}$ and $L$ previously reported within the TF 
approximation using different functionals (including relativistic ones)~\cite{Bao&Shen15,Oyamatsu&Iida07,Grill+13}. However, such a correlation does not seem to be present in calculations based on liquid-drop models~\cite{Newton+13,Balliet+21,DinhThi+21b,Parmar+22}. This could stem from the fact that predictions are very sensitive to the parametrizations of surface and curvature terms.

In addition, the correlation 
between $\widetilde{L}$ and $\bar n_\mathrm{sp}$ should not be taken at face value since $\widetilde{L}$ is defined 
at saturation density $n_0$ whereas pasta phases emerge at much lower densities down to about $n_0/4$. 
The behavior of the symmetry energy $S(n)$ at densities $n\sim(0.04-0.09)$\,fm$^{-3}$ is more relevant 
for understanding the formation of pasta. It is only because the behavior of $S(n)$ at such densities 
is directly related to $\widetilde{L}$ that $\bar n_\mathrm{sp}$ is correlated to this coefficient for BSk22, BSk24
and BSk25. However, comparing BSk24 and BSk26 leads to the opposite conclusion that $\bar n_\mathrm{sp}$
is \emph{anti}correlated to $\widetilde{L}$. The contradiction is only apparent: 
the slightly higher transition density $\bar n_\mathrm{sp}$ found for BSk26 can be easily interpreted from the slightly lower symmetry energy $S(n)$ at $\bar n \approx 0.05$\,fm$^{-3}$, as can be seen in Fig.~\ref{fig:symm} (actually, the slope of $S(n)$ at these densities is also larger for BSk26, since the $S(n)$ curve is altered by $\widetilde{K}_\mathrm{sym}$).
This comparison illustrates the importance of 
considering a family of consistently fitted functionals to clarify the role of 
symmetry energy. 

With increasing density, we find that the equilibrium density profile becomes 
flat
at some point (with $n_{\Lambda q}$ approaching zero), thus marking the transition to the core. It is worth noting, that allowing for pasta (especially for bucatini and Swiss cheese) slightly shifts the crust-core transition to higher densities $\bar n_\mathrm{cc}$, as seen in Table \ref{tab2} (and also in Fig.~\ref{fig:etf}), compared to the ones if just spherical clusters were considered, $\bar n^\mathrm{s}_\mathrm{cc}$. Besides, it is interesting to juxtapose $\bar n_\mathrm{cc}$ with the approximate value $\bar n^{*}_\mathrm{cc}$ given in Ref.~\cite{Pearson_ea18_bsk22-26} and determined from the core side as the point at which homogeneous matter becomes unstable against small spatial density fluctuations~\cite{Ducoin+07}. 
As seen from Table \ref{tab2}, the two definitions are in good agreement; the largest difference amounts to $~\sim 3\%$ for BSk25. Figure~\ref{fig:symm} and Table \ref{tab2} also show that a higher symmetry energy $S(n)$ at densities around $n_0/2$  
(lower $\widetilde{L}$) leads to a higher crust-core transition density $\bar n_\mathrm{cc}$, in agreement 
with previous studies (see, e.g., Refs.~\cite{Ducoin+10_Symm, Carreau+19} and references therein).

 \begin{table}
 	\centering
 	\caption{Average baryon number densities for the onset of pasta phases and for the crust-core transition for BSk functionals. The densities $\bar n_\mathrm{sp}$ and $\bar n^{\mathrm{SI}}_\mathrm{sp}$ marking the transition from spherical clusters to pasta which were determined from ETF and ETFSI calculations respectively. The density $\bar n_\mathrm{cc}$ ($\bar n^\mathrm{s}_\mathrm{cc}$) delimiting the crust-core boundary with (without) pasta is compared to the one, denoted here $\bar n^\mathrm{*}_\mathrm{cc}$, obtained from instabilities of homogeneous matter given in Ref.~\cite{Pearson_ea18_bsk22-26}.} 
 \label{tab2}
 \begin{tabular} {cccccc}
 	\hline 
 	&$\bar n_\mathrm{sp}$\,[fm$^{-3}$]&$\bar n^{\mathrm{SI}}_\mathrm{sp}$\,[fm$^{-3}$]&$\bar n^\mathrm{s}_\mathrm{cc}$\,[fm$^{-3}$]&$\bar n_\mathrm{cc}$\,[fm$^{-3}$]&$\bar n^\mathrm{*}_\mathrm{cc}$\,[fm$^{-3}$]\\
	\hline
 	BSk22&0.056&0.066&0.071&0.071&0.072 \\
 	BSk24&0.050&0.076&0.081&0.082&0.081 \\
 	BSk25&0.042&0.065&0.086&0.089&0.086 \\
 	BSk26&0.056&0.079&0.085&0.086&0.085 \\
 	\hline
 \end{tabular}
\end{table}

All in all, functionals with higher $S(n)$ at densities relevant for neutron-star crusts (generally 
corresponding to the lower slope of the symmetry energy) yield larger density regions filled by pasta phases in the ETF approach.   

\subsection{Inclusion of microscopic corrections}
 \label{sec:etfsi}

\begin{figure}
 	\includegraphics[width=0.7\columnwidth,angle=-0]{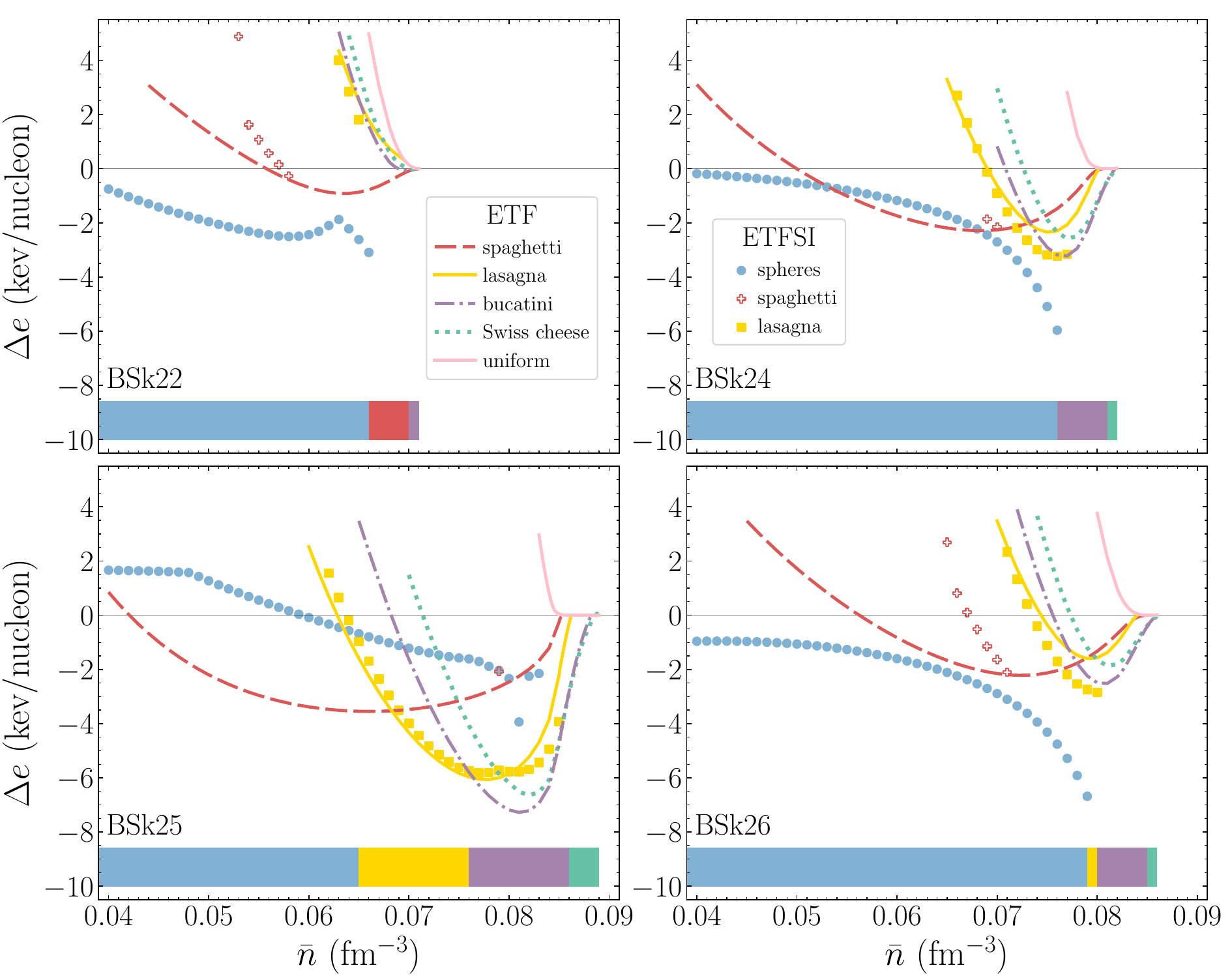}
 	\caption{ETFSI and ETF energy per nucleon (in MeV) for different pasta shapes with ETF energy of spheres subtracted versus average baryon number density $\bar n$ (in fm$^{-3}$). ETF results are plotted by curves with the same legend as in Figure~\ref{fig:etf}. ETFSI results are displayed by symbols. Solid blue circles and yellow squares correspond to solutions with bound protons for spheres and for lasagna respectively, while empty red crosses to solutions for spaghetti with dripped protons (see text for details). The horizontal histograms show the sequence of pasta phases with microscopic corrections.}
 	\label{fig:etfsi}
 \end{figure}

In the ETFSI approach, we first determine the geometric parameters for each phase at the ETF level before including microscopic corrections by fixing the value of $Z$ corresponding to the proton number in the WS cell in the case of spherical structures, the number per unit length in the case of 
cylindrical structures, and the number per unit area in the case of lasagna. 
Then we calculate the ETFSI energy per nucleon~\eqref{eq:eetfsi} and vary $Z$ to find the optimum configuration for each type of pasta. Since the microscopic corrections can significantly shift the optimum value of Z, the minimization should be performed over a sufficiently wide range of Z values. Moreover, as was mentioned in our previous work~\cite{Pearson+22}, our approach implies dropping the SI and pairing corrections beyond the onset of proton drip, defined by the point at which some protons become ``unbound'' in the sense that the energy of the last occupied proton single-particle state exceeds the value of the proton potential at the border of the cell. Consequently, no corrections are added for Swiss cheese and bucatini. For each given $\bar{n}$ the proton-drip point is characterized by $Z$. 
To start with, for spheres, we consider the range of $Z$ from 20 to 200. 
We determine the equilibrium configuration for each given average baryon density $\bar n$ by comparing the ETFSI or ETF energies per nucleon for different $Z$ values depending on whether protons are all bound or some of them are free in the sense defined above. The results are displayed as blue circles in Figure~\ref{fig:etfsi}. To better assess the importance of the microscopic corrections, the ETF energy of spheres is subtracted. We find for spheres that the optimal $Z$ value always corresponds to one with bound protons up to some density, beyond which protons are unbound for the entire interval of $Z$. In such cases,  the ETFSI energy reduces to that obtained in the ETF method, whence the absence of blue circles in Figure~\ref{fig:etfsi} at high densities.
We can see that the correction is negative for almost all densities (with the notable exception of BSk25), 
and has, therefore, a stabilizing effect for spheres. Note that spikes in the ETFSI energy for spheres are caused by changes in the optimum $Z$ value. 

For lasagna and spaghetti, the situation is a bit more delicate and we proceed as follows. We search for the minimum of the ETFSI energy per nucleon for each given baryon density $\bar n$ by varying $Z$ near the equilibrium ETF value $Z_\mathrm{ETF}$ within $\Delta Z\sim0.015$\,fm$^{-2}$ for lasagna and $\Delta Z\sim 1$\,fm$^{-1}$ for spaghetti. For the latter we find that the microscopic corrections are generally positive, and it might happen that the equilibrium ETFSI configuration actually corresponds to an ETF spaghetti solution for some $Z$ different from $Z_\mathrm{ETF}$ if proton drip occurs. 
To make sure we do not miss solutions of this kind, 
we check that the range of $Z$ considered includes all values for which the ETF energy per nucleon of spaghetti is lower than the optimum ETFSI energy per nucleon among all the other possible phases.

 Comparing Figs.~\ref{fig:etf} and \ref{fig:etfsi} shows that including or not the microscopic corrections leads to strikingly different predictions. For all functionals, the density range over which pasta phases are present is considerably reduced in the ETFSI approach, thus confirming the conclusions from our previous study~\cite{Pearson+22}: the average baryon densities marking the transition between spherical clusters and pasta, indicated in Table~\ref{tab2}, are shifted to significantly higher values. Moreover, spaghetti vanish completely for all models but BSk22. The absence of spaghetti stems from the fact that the microscopic corrections are generally negative for spheres (except for BSk25 at densities $\bar n\lesssim 0.06$~fm$^{-3}$), large and positive for spaghetti, and quite small for lasagna. 
 In fact, the corrections for spaghetti are so large that the ETFSI energies lie off scale in Fig.~\ref{fig:etfsi} when all protons are bound (empty red crosses in Fig.~\ref{fig:etfsi} represent configurations in the proton-drip regime). 
 Thus, pasta phases appear only beyond the onset of proton drip for BSk22, BSk24, and BSk26. 
 The proton condensation energy, which varies in the range of $(0.1-0.3)$~keV per nucleon, remains essentially the same for the different shapes and therefore has a negligible impact on the results.

Within the ETFSI approach, the correlation between the symmetry energy and pasta phases becomes less obvious. This is because the onset of pasta phases now depends on the occurrence of proton drip except for BSk25. The densest layers of the crust still consist of Swiss cheese and bucatini and since the microscopic corrections are dropped for these phases, the crust-core transition density and its correlation with the symmetry energy are not affected.

\subsection{Abundance of nuclear pasta}
 \label{sec:mass}

To determine the amount of nuclear pasta present in a neutron star, one has to solve the well-known Tolman-Oppenheimer-Volkoff (TOV) equations describing the general-relativistic structure of a spherical star in hydrostatic equilibrium. Since the mass of the crust and its thickness are small compared to the total gravitational mass $\Mns$ and radius $\Rns$ of the star, the TOV equations can be approximately recast into a Newtonian form as (see, e.g., Ref.~\cite{ch08})
\begin{equation}\label{eq:Newton}
	\frac{{\rm d}P}{{\rm d}z} \approx g_s \rho 
\end{equation}
where $P$ is the pressure, $z$ is the proper depth below the surface, and with the surface gravity ($r_g=2 G\Mns/c^2$ is the Schwarzschild radius). 
\begin{equation}
	g_s = \frac{G\Mns}{\Rns^2}\left(1 - \frac{r_g}{\Rns}\right)^{-1/2}\, .
\end{equation}
Within this approximation, the baryonic mass contained in a layer spanning the range of pressures $\delta P$ is approximately given by~\cite{Pearson+11,Chamel20} 
\begin{equation}\label{eq:Mlayer}
	\delta M \approx \frac{4\pi \Rns^2}{g_s} \left(1-\frac{r_g}{\Rns}\right)\delta P \, .
\end{equation}
The total baryonic mass $M_{\rm crust}$ of the crust is obtained by setting $\delta P=P_{\rm cc}$, where $P_\mathrm{cc}$ stands for the pressure at the crust-core interface. The relative abundances of the different pasta phases are thus found to be independent of the global structure of the star and are given by 
\beqy\label{eq:Mpasta}
\frac{\delta M}{M_\mathrm{crust}}\approx \frac{\delta P}{P_\mathrm{cc}}\quad .
\eeqy
As shown by Equations (B25), (B28) in the appendix of Ref.~\cite{Pearson+12}, the pressure of any given ETFSI configuration can be 
calculated by 
adding to the pressure of an ideal electron Fermi gas, the pressure of homogeneous nuclear 
matter at the background nucleon densities $n_{Bq}$ in the Wigner-Seitz. 

\begin{figure}
 	\includegraphics[width=0.6\columnwidth,angle=-0]{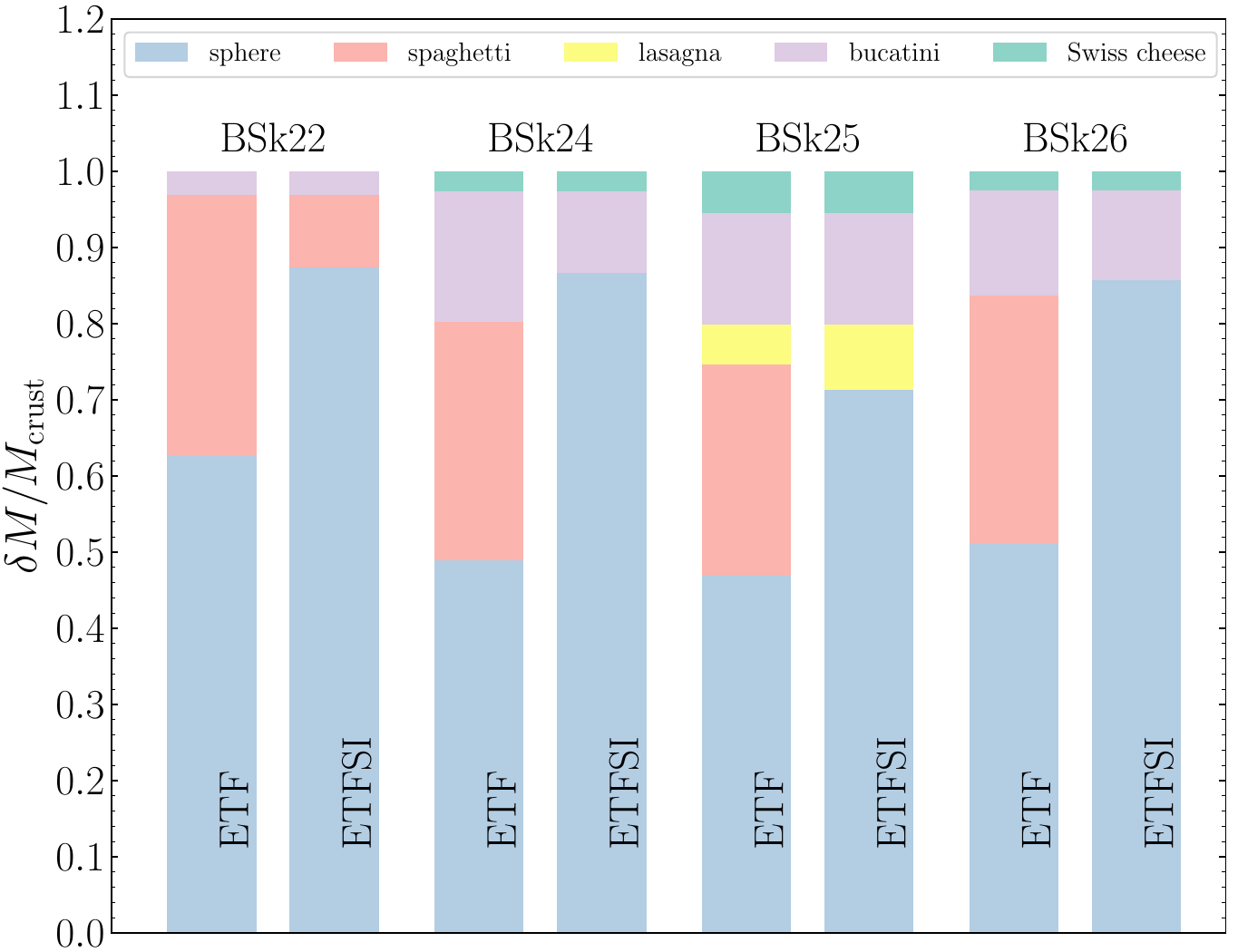}
 	\caption{Relative abundances of pasta phases in the crust of neutron stars for four nuclear functionals within the ETF and ETFSI approaches. Blue regions represent spheres, red correspond to spaghetti, yellow are for lasagna,  purple indicate bucatini and finally green are for Swiss cheese. }
 	\label{fig:mass}
 \end{figure}

Results for the different functionals are summarized in Figure~\ref{fig:mass}. 
Within the ETF method, pasta phases account for about $35\%-55\%$ of the mass of the crust. These values are in good agreement with those obtained from systematic analyses of liquid-drop models \cite{DinhThi+21a,Balliet+21,Parmar+22,Newton+13}. Taking into account the microscopic corrections leads to a drastic reduction of pasta abundance, down to about $15\%$
for BSk22, BSk24, and BSk26. For BSk25, the pasta region shrinks substantially but still represents about $30\%$ of the mass of the crust. Because the pressure must vary continuously inside a neutron star, the phase transition 
pressures are determined using the Maxwell construction. For BSk26, the resulting density jumps in ETFSI exclude lasagna. 

The different mass fractions of pasta phases predicted by the different functionals cannot be 
fully understood from the symmetry energy alone. Indeed, ignoring the 
contribution from electrons and protons (only beyond proton drip), the pressure can be 
expressed in terms of the symmetry energy as 
\begin{equation}
P = n_{Bn}^2  \frac{\partial e(n,0)}{\partial n}\biggr\vert_{n=n_{Bn}} + n_{Bn}^2 \frac{dS(n))}{dn}\biggr\vert_{n=n_{Bn}}\, . 
\end{equation}
It turns out that the two terms are of the same order but of opposite signs, recalling that $n_{Bn} \leq \bar n_\mathrm{cc}\approx n_0/2$. 
Furthermore, even using the expansion~\eqref{eq:e-expansion} with $\eta\approx 1$, it can be seen 
that the pressure is approximately given by 
\begin{equation}
P\approx \frac{n_{{\rm B}n}^2}{n_0}\left[\frac{1}{3}L + 
\frac{1}{9}(K_v + K_\mathrm{sym})\frac{n_{{\rm B}n}-n_0}{n_0}\right] \quad ,
\end{equation}
and therefore is not simply correlated to $L$, as is generally stated, but 
depends also on $K_v$, $K_\mathrm{sym}$, and $n_{{\rm B}n}$.  As the mass fractions~\eqref{eq:Mpasta} involve various transition pressures associated with different values for $n_{{\rm B}n}$, they depend on the symmetry energy in a complicated way.

\section{Conclusions}
\label{sec:concl}

In this work, we have extended our previous studies~\cite{Pearson_ea18_bsk22-26,Pearson+20,Pearson+22} of the equation of state of cold non-accreting neutron stars, focusing here on the role of the symmetry energy in the appearance of nuclear pasta phases within the ETF and ETFSI approaches. To this end, we have considered  the series of Brussels-Montreal functionals BSk22, BSk24, BSk25, and BSk26, all being accurately fitted to nuclear masses but imposing various symmetry-energy coefficients $J$ (BSk22, BSk24, BSk25) or constraining to reproduce a different neutron-matter equation of state (BSk26). These functionals predict 
different behaviors for the symmetry energy at subsaturation densities relevant to the pasta phases.

Examining first the ETF results, pasta phases consisting mostly of spaghetti are present for all employed functionals. 
Applying a new computer code that allows for a more accurate and reliable investigation of 
the bottom layers of the crust, the region previously found to be made of lasagna for BSk24 
is now replaced by bucatini and Swiss cheese. Lasagna is actually absent for all models but 
BSk25, thus for them the sequence of pasta phases does not follow the general predictions from simple liquid-drop models~\cite{Hashimoto+84}. 
However, the role of symmetry energy is found to be similar to that observed in previous studies based on 
the TF method~\cite{Bao&Shen15,Oyamatsu&Iida07,Grill+13}, namely, pasta phases are more likely to 
appear for models with higher values of the symmetry energy at the relevant densities 
(corresponding to lower values for $\widetilde{J}$ and $\widetilde{L}$): the threshold 
density $\bar n_\mathrm{sp}$ for the onset of pasta phases is lower while the crust-core 
transition density $\bar n_\mathrm{cc}$ is higher. We find that pasta phases represent a sizeable 
fraction, $35\%-55\%$, of the mass of neutron-star crusts in agreement with previous estimates
from liquid-drop treatments~\cite{Newton+13,Balliet+21,DinhThi+21a,Parmar+22}.

However, when we add microscopic corrections consistently on top of ETF via the ETFSI approach, 
the region containing pasta shrinks dramatically. In particular, the spaghetti phase disappears 
for BSk24, as found earlier~\cite{Pearson+22}. The spaghetti phase also vanishes with BSk25 and BSk26. 
Although it is still present for BSk22, it occupies a much narrower domain. As a consequence, the threshold 
density $\bar n_\mathrm{sp}$ for the onset of pasta is shifted to much higher values. We no longer find 
clear correlations with the symmetry energy, whose influence appears to be overshadowed by the microscopic corrections. The pasta phases in the densest layers of the crust and the crust-core transition density 
$\bar n_\mathrm{cc}$ remain unchanged. Therefore, the overall abundance of pasta drops down to 
$\approx 15-30\%$ only. This shows the importance of microscopic effects for determining the structure of 
the nuclear pasta mantle.  Nevertheless, our conclusions need to be confirmed by fully self-consistent 
mean-field calculations.

\begin{acknowledgments}
 N.S. is thankful to M.E. Gusakov and A.I. Chugunov for useful comments on the early developments of the code and to S. Goriely for discussions. This work was financially supported by Fonds de la Recherche Scientifique (Belgium) under Grant No. IISN 4.4502.19. It has also received funding from the FWO (Belgium) and the Fonds de la Recherche Scientifique (Belgium) under the Excellence of Science (EOS) programme (project No. 40007501).
 
\end{acknowledgments}

\bibliography{literature}

\end{document}